\documentclass[onecolumn,showpacs,preprintnumbers,amsmath,amssymb]{revtex4}

\usepackage{graphicx}
\usepackage{dcolumn}
\usepackage{bm}

\begin{document}

\preprint{APS/123-QED}

\title{Hydrogen Atom in the Cosmic Microwave Background}

\author{Jose A. Magpantay}

\email{jose.magpantay11@gmail.com}
\affiliation{Quezon City, Philippines}%
 
\date{\today}

\begin{abstract}
The cosmic microwave background covers the entire universe which suggests the absence of any closed system, except the universe itself. In this paper, I consider the effect of the cosmic microwave background on the hydrogen atom, which must be very small otherwise changes in energy levels would have been measurable. But how small is small? This I compute by considering a system in an environment or bath. I derive the bath's (in this case the CMB) effect on the hydrogen atom in the Feynman-Vernon approach to an open system. The effect is small and quantified in terms of a correction to the hydrogen atom that breaks time-reversal symmetry, as expected of memory effects. This is significant, there are imperceptible changes in the state of the hydrogen atom, which means that the pervasive CMB must have similar small effect on other atoms thus breaking time-reversal symmetry in all physical systems. 
\end{abstract}

\pacs{Valid PACS appear here}
\maketitle


1. Quantum theory started from two outstanding problems in the early 20th century - the blackbody radiation pattern and the spectral lines of the hydrogen atom - see any introductory atomic physics book, for example \cite{Eisberg}. In this paper, I consider the effect of the cosmic microwave background on the hydrogen atom, which we know from cosmology started at 380,000 years after the Big Bang when the Universe has cooled to temperature of the order of $ 10^{3} $ K and has expanded to the size of the order of $ 10^{7} $ light years, reducing the energy of the electromagnetic radiation to the order of $ 10^{-1} $ ev \cite{Wikipedia1}, which is too little to kick out the electron from the clutches of the proton (ionization energy is of the order of ev). At 13.8 B years of age, the universe is bathed by a microwave background radiation of cosmic origin, the electromagnetic radiation that became transparent to the universe at 380,000 years age, now has a much lower energy of the order of $ 10^{-3} $ ev. The microwave background follows extremely well the blackbody radiation pattern derived by Planck at the turn of the century with temperature equal to 2.73 K \cite{Wikipedia2}. About a decade later, Bohr explained the spectral lines of hydrogen atom by considering an isolated proton and electron system by assuming classical physics - planetary orbits, but proposing quantized angular momentum states. Still another decade later, the hydrogen atom was treated using the new quantum mechanics formulated by Heisenberg and Schroedinger, spin was later added and the electron made relativistic by Dirac, showing a richer structure with corresponding spectral lines, see for example \cite{Schiff}. 

But up to this time, the treatment of the hydrogen atom neglects the cosmic microwave background. Although each individual photon does not have enough energy to alter substantially the energy of the electron, however, there are around $ 10^{2} $ per $ cm^{3} $ of CMB photons in the universe today, every electron and proton must then be hit by CMB photons every now and then because the number of particles in the universe today is of the order of $ 10^{-7} $ per $ cm^{3} $, thus giving around $ 10^{9} $ CMB photons per hydrogen atom \cite{Wikipedia1}. This suggests a situation of a Brownian particle in a very weak and dilute (typically water has $ 10^{22} $ molecules per $ cm^{3} $ that can hit a pollen) 'fluid'. 

The question now is how to compute the effect of the CMB on the hydrogen atom? The Brownian particle analogy is not too useful. Fortunately, the quantum formalism for dealing with any system in bath or environment, called an open system, is established. The density matrix of such a system, after taking care of the bath degrees of freedom is given by the Feynman/Vernon method \cite{Feynman} in non-relativistic quantum case. Thus, I will I only consider the non-relativistic hydrogen atom. The fully relativistic quantum theory of electron and photon is the realm of QED.

2. The CMB is the remnant of the EM radiation that decoupled from matter at 380 K after the Big Bang. It is described then by a free EM field with Lagrangian (SI units)
\begin{equation}\label{1}
S_{em} = -\frac{1}{4\mu_{0}} \int d^{4}x (\partial_{\mu} A_{\nu} - \partial_{\nu} A_{\mu})(\partial^{\mu} A^{\nu} - \partial^{\nu} A^{\mu}),
\end{equation}
where $ \mu_{0} $ is the magnetic permeability of free space. I use the metric with signature (-+++). The use of the Coulomb gauge $ \vec{\partial}\cdot\vec{A} = 0 $ results in 
\begin{equation}\label{2}
S_{em} = \frac{1}{2\mu_{0}} \int d^{4}x \left[(\partial_{0}A_{i})^{2} - (\partial_{i}A_{j})^{2} \right] ,
\end{equation}
giving the field equation
\begin{equation}\label{3}
(\partial_{0}^{2} - \vec{\partial}\cdot\vec{\partial})\vec{A} = 0.
\end{equation}
The solution to equation (3) is given by 
\begin{subequations}\label{4}
\begin{gather}
\vec{A}(\vec{x},t) = \sum_{\vec{k}, \alpha} \left[ q_{\vec{k}}(t) \vec{\varphi}_{\vec{k}, \alpha}(\vec{x}) +q_{\vec{k}}(t) \vec{\varphi}^{*}_{\vec{k},\alpha}(\vec{x}) \right],\label{first}\\
\vec{\varphi}_{\vec{k} \alpha}(\vec{x}) = (\frac{1}{2\pi})^{\frac{3}{2}} \vec{\epsilon}(\vec{k},\alpha) \exp{(i\vec{k}\cdot\vec{x})},
\end{gather}
\end{subequations}
where $ \vec{\epsilon}(\vec{k},\alpha) $ with $ \alpha = 1, 2 $ are the polarization vectors of unit length and $ ( \vec{\epsilon}(\vec{k},1), \vec{\epsilon}(\vec{k},2), \vec{k} ) $ form a right-handed triad of vectors. Also, the $ \vec{k} $ summation will go into continuum integration when the CMB nature of the EM waves is accounted for later.
For equation (4) to be a solution to equation (3), the $ q_{\vec{k}}(t) $ must satisfy
\begin{equation}\label{5}
\dfrac{d^{2} q_{\vec{k}}}{dt^{2}} + (ck)^{2} q_{\vec{k}} = 0,
\end{equation}
which means $ q_{\vec{k}} $ is a harmonic oscillator of 'mass' $ \frac{1}{\mu_{0}c} $ and 'frequency' ck. It is assumed in equation (4) that the harmonic oscillator solutions are all real (sine and cosine solutions will be used). Substituting equations(4;a,b) in equation (2) results in
\begin{equation}\label{6}
S_{em} = \int dt \sum_{\vec{k}} \frac{1}{2\mu_{0} c}\left[ \left(\dfrac{dq_{\vec{k}}}{dt}\right)^{2} - (ck)^{2} q_{\vec{k}}^{2} \right].
\end{equation}
In this form, the free EM wave is written as a sum of harmonic oscillators.

3. A charge particle of charge q interacts with an electromagnetic wave as given in the interaction Lagrangian term
\begin{equation}\label{7}
L_{int} = q\vec{v} \cdot \vec{A}(\vec{x},t),
\end{equation}
A hydrogen atom in a CMB has both proton and electron interacting with the CMB EM wave. Since the proton has much smaller velocity than the electron in the hydrogen atom (typically of the order of $ 10^{-4} $ slower), I will neglect the CMB interaction with the proton and just consider the effect on the electron. For simplicity, I will just put the proton at the origin. The full Lagrangian being considered is now
\begin{subequations}\label{8}
\begin{gather}
L = L_{hy} + L_{em} + L_{int},\label{first}\\
L_{hy} = \frac{1}{2} m (\dfrac{d\vec{x}}{dt})^{2} + \dfrac{e^{2}}{4\pi \epsilon_{0}} \frac{1}{x}, \label{second} \\
L_{em} = \sum_{\vec{k}} \frac{1}{2\mu_{0}c}\left[(\dfrac{dq_{\vec{k}}}{dt})^{2} - (ck)^{2} q_{\vec{k}}^{2} \right], \label{third}\\ 
L_{int} = -e \dot{\vec{x}} \cdot \sum_{\vec{k}, \alpha} \left[ q_{\vec{k}}(t) \vec{\varphi}_{\vec{k} \alpha}(\vec{x}) +q_{\vec{k}}(t) \vec{\varphi}^{*}_{\vec{k} \alpha}(\vec{x}) \right],
\end{gather}
\end{subequations}
where e is the electron charge. Using equation (4b), the interaction of the electron with the EM waves can be written as
\begin{equation}\label{9}
L_{int} = -e (\frac{1}{(2\pi})^{\frac{3}{2}} \sum_{\vec{k}} \dot{\vec{x}}(t) \cdot \vec{\epsilon} \cos (\vec{k} \cdot \vec{x}) q_{\vec{k}}(t),
\end{equation} 
where  $ \vec{\epsilon} = [\vec{\epsilon}(\vec{k},1) + \vec{\epsilon}(\vec{k},2)] $. Equations (8b) and (9) show the similarity with the point particle in a bath of harmonic oscillator discussed in F/V and Caldeira-Leggett (C/L) \cite{Caldeira} with a crucial difference of the electron having a bit more complicated interaction - it is velocity dependent and the coordinate is not factored out of the 'frequency' term as shown in the cosine. But still for every $ \vec{k} $, the calculations done by F/V and C/L are applicable.

4. I give the results of the F/V and C/L calculations right up to the point before summing over the different 'frequency' contributions, which I will do in 5. But before I go into that, an admission of error. In a previous paper \cite{Magpantay}, I argued the difference between my calculation's equation (25) and C/L's equation (3.9) and (3.10). First, I missed a factor of $ \frac{1}{2} $ in the second line of equation (25) and second, I missed a simple evaluation that would have shown $ (\coth \theta + \dfrac{1}{\sinh \theta} ) = \coth \frac{\theta}{2} ) $. Taking these corrections, I find the C/L results and there is no basis for claiming any difference with their result. 

The C/L quantities, as given in formulas (3.9) to (3.11), translate to this paper's quantities via
\begin{subequations}\label{10}
\begin{gather}
m \rightarrow \frac{1}{\mu_{0}c},\label{first}\\
\omega_{i} \rightarrow ck,\label{second}\\
c_{i} \rightarrow e (\frac{1}{(2\pi})^{\frac{3}{2}}, \label{third} \\
x \rightarrow \dot{\vec{x}}(t) \cdot \vec{\epsilon} \cos (\vec{k} \cdot \vec{x}).
\end{gather}
\end{subequations} 
The propagator for the density matrix is given by
\begin{equation}\label{11}
\begin{split}
J(\vec{x},\vec{y};\vec{x}',\vec{y}')& = \int_{end points} (d\tilde{\vec{x}}) (d\tilde{\vec{y}}) \exp \frac{i}{\hbar} \big\lbrace \int_{0}^{t} dt' \left[ L_{hy}(\tilde{\vec{x}}) - L_{hy}(\tilde{\vec{y}}) \right] \\
& \quad + \int_{0}^{t} dt' \int_{0}^{t'} dt'' \sum_{\vec{k}} \frac{\mu_{0}e^{2}}{16 \pi^{3}} \frac{1}{k}\Big\lbrace \left[ \dot{\tilde{\vec{x}}}(t') \cdot \vec{\epsilon} \cos (\vec{k} \cdot \tilde{\vec{x}}(t')) - \dot{\tilde{\vec{y}}}(t') \cdot \vec{\epsilon} \cos (\vec{k} \cdot \tilde{\vec{y}}(t') \right] \\
& \quad X i \coth (\frac{\beta \hbar ck}{2}) \cos (ck(t'-t'')) \left[ \dot{\tilde{\vec{x}}}(t'') \cdot \vec{\epsilon} \cos (\vec{k} \cdot \tilde{\vec{x}}(t'')) - \dot{\tilde{\vec{y}}}(t'') \cdot \vec{\epsilon} \cos (\vec{k} \cdot \tilde{\vec{y}}(t'')) \right]\\
& \quad + \Big[ \dot{\tilde{\vec{x}}}(t') \cdot \vec{\epsilon} \cos (\vec{k} \cdot \tilde{\vec{x}}(t')) - \dot{\tilde{\vec{y}}}(t') \cdot \vec{\epsilon} \cos (\vec{k} \cdot \tilde{\vec{y}}(t') \Big] \sin (ck(t'-t'')) \\
& \quad X\Big[ \dot{\tilde{\vec{x}}}(t'') \cdot \vec{\epsilon} \cos (\vec{k} \cdot \tilde{\vec{x}}(t'')) + \dot{\tilde{\vec{y}}}(t'') \cdot \vec{\epsilon} \cos (\vec{k} \cdot \tilde{\vec{y}}(t'')) \Big\rbrace \big\rbrace .
\end{split}
\end{equation}
This propagator is more involved because when we sum over $ \vec{k} $, the electron coordinate $ \vec{x} $ is entwined with it inside the cosine term and the electron velocity is entwined with the sum of the two polarization vectors.  But the more important question is what weighs the contribution of the different $ \vec{k} $?

5. Since the electromagnetic waves are supposed to represent the CMB, the sum over $ \vec{k} $ will naturally go to the integral
\begin{equation}\label{12} 
\sum_{\vec{k}} \rightarrow \int k^{2} dk d\Omega (1)
\end{equation}
which means all 'frequencies'  have equal weight because of the weight (1), i.e., the EM waves represent a 'white' light (note, the $ k^{2} $ and $ d\Omega $ are geometrical factors in 3D). This is not representative of CMB. To ensure that the EM waves in equation (11) is a CMB, it must be weighted on by the blackbody radiation pattern given by Planck. 
The Planck distribution given by
\begin{equation}\label{13}
\rho_{T}(\lambda) d\lambda = \frac{8\pi h c}{\lambda^{5}} \dfrac{1}{ \exp {\frac{\beta h c}{\lambda}} - 1} d\lambda,
\end{equation}
where $ \beta = \frac{1}{k_{B}T} $, $ T = 2.7 K $ and $ k_{B} $ is Boltzmann's constant. The distribution accounts for the number of waves and the average energy per wave between wavelengths $ \lambda $ and $ \lambda + d\lambda $. Since the temperature is 2.7 K, I am not considering time starting from 380 K years after the Big Bang (temperature then is of the order of $ 10^{3} $ K but only when the Universe has cooled down to 2.7 K. Current ideas do not know when this exactly happened but present ideas speculate that the Universe has cooled down to 10 to 20 K after a billion years so maybe a few more billion years will cool down the Universe to about 2.7 K. 

In terms of the 'frequency' k, which is actually the wave number $ k = \frac{2\pi}{\lambda} $, the Planck distribution is given by
\begin{equation}\label{14}
\rho_{T}(\lambda) d\lambda = - \rho_{T}(k) dk,
\end{equation}
giving
\begin{equation}\label{15}
\rho_{T}(k) = \frac{h c}{2 \pi^{3}} k^{3} \dfrac{1}{ \exp {\beta \hbar c k} - 1}.
\end{equation}
To have a normalized distribution, so that the contribution of each 'frequency' k is properly weighted, the distribution $ \rho_{T}(k) $ must be divided by the norm 
\begin{equation}\label{16}
\begin{split}
\rho_{T}& = \int_{0}^{\infty} dk \rho_{T}(k)\\
			 & =  (\dfrac{2\pi}{\beta \hbar c})^{4} \frac{1}{240} 
\end{split}
\end{equation}
where the integral is given in page 325 of \cite{Gradshteyn}.

The sum over $ \vec{k} $ in equation(11) should now be replaced by
\begin{equation}\label{17}
\begin{split}
\sum_{\vec{k}} (eqn 11 term) & = \int d\Omega \int_{0}^{\infty} dk OF \dfrac{1}{\exp {\beta \hbar c k} - 1} k^{4} \big\lbrace \Big[ \dot{\tilde{\vec{x}}}(t') \cdot \vec{\epsilon} \cos (\vec{k} \cdot \tilde{\vec{x}}(t')) - \dot{\tilde{\vec{y}}}(t') \cdot \vec{\epsilon} \cos (\vec{k} \cdot \tilde{\vec{y}}(t') \Big] \\
& \quad X i \coth (\frac{\beta \hbar ck}{2}) \cos (ck(t'-t'')) \Big[ \dot{\tilde{\vec{x}}}(t'') \cdot \vec{\epsilon} \cos (\vec{k} \cdot \tilde{\vec{x}}(t'')) - \dot{\tilde{\vec{y}}}(t'') \cdot \vec{\epsilon} \cos (\vec{k} \cdot \tilde{\vec{y}}(t'')) \Big]\\
& \quad + \Big[ \dot{\tilde{\vec{x}}}(t') \cdot \vec{\epsilon} \cos (\vec{k} \cdot \tilde{\vec{x}}(t')) - \dot{\tilde{\vec{y}}}(t') \cdot \vec{\epsilon} \cos (\vec{k} \cdot \tilde{\vec{y}}(t') \Big] \sin (ck(t'-t'')) \\
& \quad X\Big[ \dot{\tilde{\vec{x}}}(t'') \cdot \vec{\epsilon} \cos (\vec{k} \cdot \tilde{\vec{x}}(t'')) + \dot{\tilde{\vec{y}}}(t'') \cdot \vec{\epsilon} \cos (\vec{k} \cdot \tilde{\vec{y}}(t'')) \big\rbrace,
\end{split}
\end{equation}
where the constant overall factor OF is given by
\begin{equation}\label{18}
\begin{split}
OF & = e^{2} \mu_{0} h c (\beta \hbar c)^{4} \frac{240}{256 \pi^{9}} \\
	  & \propto 10^{-82}.
\end{split}
\end{equation}
The SI values of the physical constants and $ T = 2.7 $ K are used in above. Note, the overall factor is fantastically small. It gives a sense that the CMB memory correction to the action must really be small.
 	  	  
6. The above integral over k is rather complicated and the focus of this part. To make the integration doable, define
\begin{equation}\label{19}
u = ck.
\end{equation} 
Using this in the $ \coth (\frac{\beta \hbar ck}{2}) $ term, the term can be approximated by 
\begin{equation}\label{20}
\coth (\frac{\beta \hbar ck}{2}) \approx \frac{2}{\beta \hbar}(\frac{1}{u}).
\end{equation} 
The reason for this is that substituting the values for the Boltzmann constant $ k_{B} $ and Planck constant $ \hbar $, the RHS of (20) is $ \propto 10^{11} \frac{1}{u} $ and the neglected terms are of the order of $ (10^{-11} u)^{n} $, where $ n = 1, 2,... $. 

Another term in equation (17) is $ \cos (\vec{k} \cdot \tilde{\vec{x}}(t')) = \cos ( \frac{1}{c} x(t') u \cos \Theta_{x}(t') ) $, where $ \Theta_{x}(t')  $ is the angle between the electron's position vector at t' and the CMB's wave number vector $ \vec{k} $, with an explicit time dependence as shown later. Since $ \frac{1}{c} \approx 10^{-8} $, then the approximation of the cosine term can be used, i.e., 
\begin{equation}\label{21}
\cos ( \frac{1}{c} x(t') u \cos \Theta_{x}(t') ) \approx 1 - \frac{1}{2} (\dfrac{{x}(t') \cos \Theta_{x}(t') u}{c})^{2}.
\end{equation}

Another angle that appears in equation (17) is in the term $ \dot{\tilde{\vec{x}}}(t') \cdot \vec{\epsilon} = \vert\dot{\tilde{\vec{x}}}(t')\vert \cos \Phi_{x}(t') $, the angle between the electron's velocity vector at t' and the CMB's polarization vector $ \vec{\epsilon} $. This is also time dependent and the explicit time dependence will also be shown later.

Using equations (19), (20) and (21) in (17) results in
\begin{equation}\label{22}
\begin{split}
Equation (17)& = ( 10^{-122}) \int d\Omega \int_{0}^{\infty} du \dfrac{u^{4}}{ \exp {\gamma u} - 1} \big\lbrace i\dfrac{2}{\beta \hbar} \dfrac{\cos u(t'-t'')}{u} \\
& \quad X \left[ \vert\dot{\tilde{\vec{x}}}(t')\vert \cos \Phi_{x}(t') \left( 1 - \frac{1}{2} (\dfrac{x(t') \cos \Theta_{x}(t') u}{c})^{2} \right) - \vert\dot{\tilde{\vec{y}}}(t')\vert \cos \Phi_{y}(t') \left( 1 - \frac{1}{2} (\dfrac{y(t')\cos \Theta_{y}(t') u}{c})^{2} \right) \right] \\  
& \quad X \left[ \vert\dot{\tilde{\vec{x}}}(t'')\vert \cos \Phi_{x}(t'') \left( 1 - \frac{1}{2} (\dfrac{x(t'') \cos \Theta_{x}(t'') u}{c})^{2} \right) - \vert\dot{\tilde{\vec{y}}}(t'')\vert \cos \Phi_{y}(t'') \left( 1 - \frac{1}{2} (\dfrac{y(t'') \cos \Theta_{y}(t'') u}{c})^{2} \right) \right] \\
& \quad + \sin u(t'-t'') \left[ \vert\dot{\tilde{\vec{x}}}(t')\vert \cos \Phi_{x}(t') \left( 1 - \frac{1}{2} (\dfrac{x(t') \cos \Theta_{x}(t') u}{c})^{2} \right) - \vert\dot{\tilde{\vec{y}}}(t')\vert \cos \Phi_{y}(t') \left( 1 - \frac{1}{2} (\dfrac{y(t') \cos \Theta_{y}(t') u}{c})^{2} \right) \right] \\
& \quad X \left[ \vert\dot{\tilde{\vec{x}}}(t'')\vert \cos \Phi_{x}(t'') \left( 1 - \frac{1}{2} (\dfrac{x(t'') \cos \Theta_{x}(t'') u}{c})^{2} \right) + \vert\dot{\tilde{\vec{y}}}(t'')\vert \cos \Phi_{y}(t'') \left( 1 - \frac{1}{2} (\dfrac{y(t'') \cos \Theta_{y}(t'') u}{c})^{2} \right) \right] \big\rbrace,
\end{split}
\end{equation} 
where $ \gamma = \dfrac{\hbar}{k_{B}T} $ and $ T = 2.7 K $. Note the considerable reduction of the OF by $ \frac{1}{c^{5}} \propto 10^{-40} $. 

From the Table of Integrals \cite{Gradshteyn}
\begin{subequations}\label{23}
\begin{gather}
\int_{0}^{\infty} \dfrac{\sin a u}{\exp {\gamma u} - 1} du = \frac{\pi}{2\gamma} \coth (\frac{ \pi a}{\gamma}) - \frac{1}{2a},  [ a > 0, Re \gamma > 0 ] \label{first} \\
\int_{0}^{\infty} \dfrac{u \cos bu}{\exp {\gamma u} - 1} du = \frac{1}{2b^{2}} - \frac{\pi^{2}}{2 \gamma^{2}} (\sinh (\frac{b\pi}{\gamma}))^{-2}, [ Re \gamma > 0],
\end{gather}
\end{subequations}
where $ a = b = (t' - t'') $. The u integrals in equation (22) are arrived at by differentiating equations (23) w.r.t. a or b a number of times to arrive at the power of u in each integral. The result of each integration is also simplified by taking note that $ \gamma = \frac{\hbar}{k_{B}T} \approx 10^{-11} $. With these in mind, I list down the three least small among all the small terms.
They are, in decreasing strength, and following equation (11), to include the time integrals:
\begin{subequations}\label{24}
\begin{gather}
\begin{split}
First& \propto (10^{-111}) \int_{0}^{t} dt' \int_{0}^{t'} dt'' \int d\Omega \left\lbrace \dfrac{3}{(t' - t'')^{4}} - \dfrac{8 \pi^{2}}{\gamma^{4}} \exp {(\dfrac{-2\pi (t' - t'')}{\gamma})} \right\rbrace  \\
& \quad X\left[ \vert\dot{\tilde{\vec{x}}}(t')\vert \cos \Phi_{x}(t') - \vert\dot{\tilde{\vec{y}}}(t')\vert \cos \Phi_{y}(t') \right] \left[ \vert\dot{\tilde{\vec{x}}}(t'')\vert \cos \Phi_{x}(t'') -\vert\dot{\tilde{\vec{y}}}(t'')\vert \cos \Phi_{y}(t'') \right], 
\\
Second& \propto (10^{-122}) \int_{0}^{t} dt' \int_{0}^{t'} dt'' \int d\Omega \left\lbrace \dfrac{-12}{(t' - t'')^{5}} + 4 (\frac{\pi}{\gamma})^{5} \exp {(\dfrac{-2\pi (t' - t'')}{\gamma})} \right\rbrace \\
& \quad X\left[ \vert\dot{\tilde{\vec{x}}}(t')\vert \cos \Phi_{x}(t') - \vert\dot{\tilde{\vec{y}}}(t')\vert \cos \Phi_{y}(t') \right] \left[ \vert\dot{\tilde{\vec{x}}}(t'')\vert \cos \Phi_{x}(t'') + \vert\dot{\tilde{\vec{y}}}(t'')\vert \cos \Phi_{y}(t'') \right],\label{second} \\
Third& \propto (10^{-127}) \int_{0}^{t} dt' \int_{0}^{t'} dt'' \int d\Omega \left\lbrace \dfrac{60}{(t' - t'')^{6}} - \dfrac{32 \pi^{5}}{\gamma^{6}} \exp {(\dfrac{-2\pi (t' - t'')}{\gamma})} \right\rbrace \\
& \quad \Big \lbrace \left[ \vert\dot{\tilde{\vec{x}}}(t')\vert \cos \Phi_{x}(t') - \vert\dot{\tilde{\vec{y}}}(t')\vert \cos \Phi_{y}(t') \right] \left[ \vert\dot{\tilde{\vec{x}}}(t'')\vert \cos \Phi_{x}(t'') x^{2}(t'') \cos ^{2} \Theta_{x}(t'')- \vert\dot{\tilde{\vec{y}}}(t'')\vert \cos \Phi_{y}(t'') y^{2}(t'') \cos ^{2} \Theta_{y}(t'') \right] \\
& \quad + \left[ \vert\dot{\tilde{\vec{x}}}(t')\vert \cos \Phi_{x}(t') x^{2}(t') \cos ^{2} \Theta_{x}(t') - \vert\dot{\tilde{\vec{y}}}(t')\vert \cos \Phi_{y}(t') y^{2}(t') \cos ^{2} \Theta_{y}(t')\right] \left[ \vert\dot{\tilde{\vec{x}}}(t'')\vert \cos \Phi_{x}(t'') - \vert\dot{\tilde{\vec{y}}}(t'')\vert \cos \Phi_{y}(t'') \right] \Big \rbrace.
\end{split}
\end{gather}
\end{subequations}
In the First, the OF is multiplied by $ \frac{1}{\beta \hbar} \propto 10^{11} $ giving a factor $ \propto 10^{-111} $. In the Second, there is no modifier to OF. In the Third, the OF is multiplied by $ \frac{1}{\beta \hbar} \propto 10^{11} $ then by $ \frac{1}{c^{2}} $, which comes from equation (21). resulting in the factor $ 10^{-127} $.  

7. Equation (24) gives the correction to the hydrogen atom action terms in the propagator for the density matrix as shown in equation (11). Evaluating the path integrals in (11) with the time delayed and non-quadratic terms in (24) is clearly not possible. Still, the angular integrations can be done and should be finite (see example in the next paragraphs) even for an arbitrary $ \tilde{\vec{x}} $ and $ \tilde{\vec{y}} $ and time derivatives. Equation (24) then gives memory terms with extremely small factors showing that the time-reversal symmetry violating terms are extremely small. 

Separating out the x from y contributions to have a clearly $ L_{eff}( \vec{x}, \dot{\vec{x}} ) $ does not seem to be possible also. However, the structure of the terms of equation (24) shows the separate x, y and cross terms have about the same contribution. This will be shown in this section by considering the effect of the CMB on the Bohr model. This will also make concrete how small the effect of the CMB is on the energy levels of the atom. 

The next problem is to integrate the angles of the CMB EM waves to complete the sum over CMB waves contribution. This is done by setting up the hydrogen atom geometry, as treated by Bohr, and superimposing the CMB vector $ \vec{k} $. It will be clear later in the calculations that I will only provide an estimate of the CMB correction to the Bohr energy levels, which is the same as the Schroedinger treatment without spin and relativistic corrections. 

\begin{figure}[h]
		\includegraphics[width=15 cm, height = 12 cm]{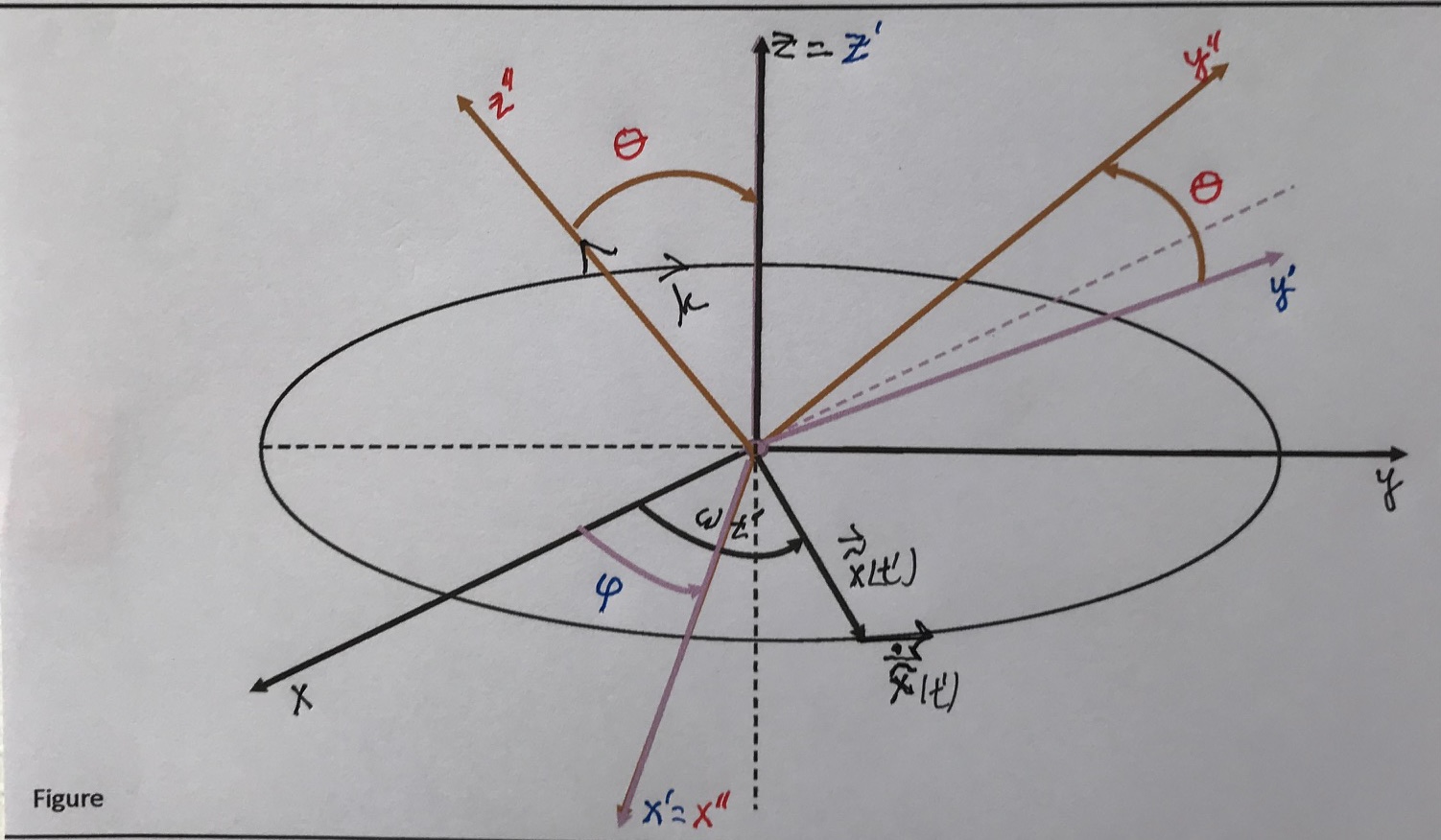}
	\caption{The electron's circular orbit on (x,y) plane superimposed with the CMB propagation vector $ \vec{k} $, which is along $ z'' $ axis and the polarization vectors along the $ x'' $ and $ y'' $ axes.}
\end{figure}
				
The electron is doing a circular motion with the Coulomb attraction from the proton providing the centripetal acceleration. The Bohr condition of the quantization of the angular momentum is then added. The relevant equations are
\begin{subequations}\label{25}
\begin{gather}
\frac{1}{4\pi \epsilon_{0}} \frac{e^{2}}{r^{2}} = m \frac{v^{2}}{r}, \label{first}\\
\begin{split}
L& = mvr \\
  & = n\hbar, 
\end{split}
\end{gather}
\end{subequations}
where n is an integer. 
These equations give the quantized values
\begin{subequations}\label{26}
\begin{gather}
r = \dfrac{4\pi \epsilon_{0} \hbar^{2}}{m e^{2}} n^{2}, \label{first}\\
v = \dfrac{e^{2}}{4\pi \epsilon_{0} \hbar} \frac{1}{n}, \label{second}\\
\omega = \dfrac{m e^{4}}{(4\pi \epsilon_{0})^{2} \hbar^{3}} \frac{1}{n^{3}}, \label{third}\\
\begin{split}
E& = - \dfrac{m e^4)}{2(4\pi \epsilon_{0})^{2} \hbar^{2}} \frac{1}{n^{2}} \\
  & = -\frac{13.6}{n^{2}} ev,
\end{split}
\end{gather}
\end{subequations}
where $ \omega $ is the angular velocity and E is the energy of the hydrogen atom.

The position and velocity of the electron, assuming it starts at $ \phi = 0 $ are given by
\begin{subequations}\label{27}
\begin{gather}
\vec{x}(t') = r \cos \omega t' \textbf{i} + r \sin \omega t' \textbf{j}, \label{first}\\
\dot{\vec{x}}(t') = - r \omega \sin \omega t' \textbf{i} + r \omega \cos \omega t' \textbf{j}.
\end{gather}
\end{subequations}
If the y terms are also considered in equation (24), the end point conditions in equation (11) shows different $ \tilde{\vec{y}} $ end points ( $ (\vec{y'}, \vec{y}) $ ). In the context of the Bohr atom, it means the y trajectory can have a different quantum number n, say n'. 
 
To specify the CMB arbitrary EM wave vector $ \vec{k} $, start with the unit vectors $ (\textbf{i}, \textbf{j}, \textbf{k}) $. Then rotate about $ \textbf{k} $ by angle $ \varphi $ to get the unit vectors $ (\textbf{i'}, \textbf{j'}, \textbf{k'}=\textbf{k}) $, then follow up with a counter clockwise rotation by angle $ \theta $ about the $ \textbf{i'} $ axis to get the unit vectors $ (\textbf{i''} = \textbf{i'}, \textbf{j''}, \textbf{k''}) $. Now I can identify these RHT of vectors with the RHT of vectors of the CMB EM wave given by $  (\vec{\epsilon}(\vec{k},1), \vec{\epsilon}(\vec{k},2), \vec{k}) $. For completeness, the vector $ \vec{k} $ makes an angle $ \theta $ with the $ \textbf{k} $ unit vector and angle $ \varphi + \frac{3\pi}{2} $ with the $ \textbf{i} $ unit vector (if I do the second rotation clockwise, the vector $ \vec{k} $ makes an angle $ \varphi + \frac{1\pi}{2} $ with the $ \textbf{i} $ unit vector . This operations will identify the terms in equation (24) as
\begin{subequations}\label{28}
\begin{gather}
\tilde{x}(t') = r, \label{first} \\
\vert\dot{\tilde{\vec{x}}}(t')\vert = r \omega, \label{second} \\
\cos \Phi_{x}(t') = \cos \omega t' (\sin \varphi + \cos \varphi \cos \theta ) - \sin \omega t' (\cos \varphi - \sin \varphi \cos \theta ), \label{third} \\
\cos \Theta_{x}(t') = \cos \omega t' \sin \varphi \sin \theta - \sin \omega t' \cos \varphi \sin \varphi.
\end{gather}
\end{subequations}

Now, the solid angle averaging can be done and all are finite integrals. The interesting part is the term
\begin{equation}\label{29}
\begin{split}
\vert\dot{\tilde{\vec{x}}}(t')\vert\vert\dot{\tilde{\vec{x}}}(t'')\vert & = v^{2}, \\
& = -\frac{2}{m} E_{n}, \\
&  \approx 10^{+31} (\frac{13.6}{n^{2}}) ev, 
\end{split}
\end{equation}
This will strengthen the really small terms given by equation (24) by 31 orders of magnitude but still resulting in very weak contributions. The change in action $ \int_{0}^{t} dt' L_{hy}(t') $ from the different contributions are now 
\begin{subequations}\label{30}
\begin{gather}
First \propto 10^{-80} \left[ \int_{0}^{t} dt' \left( \int_{0}^{t'} d\tau \left\lbrace \dfrac{3}{\tau^{4}} - \dfrac{8 \pi^{2}}{\gamma^{4}} \exp {(\dfrac{-2\pi \tau}{\gamma})} \right\rbrace \cos \omega \tau \right) \right] E_{n}, \label{first} \\
Second \propto 10^{-91} \left[ \int_{0}^{t} dt' \left( \int_{0}^{t'} d\tau \left\lbrace \dfrac{-12}{\tau^{5}} + 4 (\frac{\pi}{\gamma})^{5} \exp {(\dfrac{-2\pi \tau}{\gamma})} \right\rbrace \cos \omega \tau \right) \right] E_{n} , \label{second} \\
Third \propto 10^{-116} \left[ \int_{0}^{t} dt' \left( \int_{0}^{t'} d\tau \left\lbrace \dfrac{60}{\tau^{6}} - \dfrac{32 \pi^{5}}{\gamma^{6}} \exp {(\dfrac{-2\pi \tau}{\gamma})} \right\rbrace \cos \omega \tau \right) \right] E_{n},
\end{gather}
\end{subequations}
where a change of variables, from $ t'' $ to $ \tau = t' - t'' $ was made in the integration. The much smaller Third term comes from the fact that this term goes with $ r^{2} $, which by equation (26a), is $ \propto 10^{-20} $.

All the contributions are small, modulo the time factor given by the two time integrals, which shows memory effects. These time integrals are rapidly oscillating integrals because $ \omega \propto 10^{+14} $ by equation (26c) and modified by the damping polynomial term in the denominator and also by an exponential decaying term. The exponential damping terms are straightforward to evaluate. The problematic terms are the polynomial terms in the denominator. Still it is expected that the $ \tau $ integrations would lead to a finite integral even though the closed form in terms of $ t' $ is not known. Then the $ t' $ integration expression should give an overall effect of the CMB on the hydrogen atom, which must be small because of the very small factors that go with each term in equation (30). 

8. This paper showed that the effect of the CMB on the hydrogen atom is almost negligible, the energy levels are hardly changed, the memory effects are imperceptible but break time-reversal symmetry. Thus, this paper showed that since everything is made up of atoms, as a matter of principle, time-reversal symmetry is no longer a symmetry of atomic physics because of the CMB effect on these atoms. 

If the pervasive CMB breaks time-reversal symmetry, surely the other long-range force in the universe, gravity, can also break time-reversal symmetry. The gravitational wave generated during the Big Bang, although much weaker than the CMB effect now, must also produce time-reversal symmetry violation. In this sense, time-reversal symmetry is not really present in principle but only 'observed' because we are not sensitive to the violations.

\begin{acknowledgements}
I am grateful to Felicia Magpantay for putting my Latex file right. I would like to thank Maribel Garcia for doing the figure. A special thanks to Gravity for the constant company during this pandemic while I write this paper.
\end{acknowledgements} 

\begin{thebibliography}{8}
\bibitem{Eisberg}
Eisberg, Robert M., Fundamentals of Modern Physics, John Wiley and Sons, Inc., 1961.
\bibitem{Wikipedia1}
Wikipedia, Chronology of the Universe, last edited 09 January 2021.
\bibitem{Wikipedia2}
Wikipedia, Cosmic Microwave Background, last edited 20 January 2021.
\bibitem{Schiff}
Schiff, Leonard I., Quantum Mechanics, McGraw-Hill, Inc., 1968.
\bibitem{Feynman}
Feynman, R. P. and Vernon, F., The Theory of a General Quantum System Interacting with a Linear Dissipative System, Annals of Physics, 24, 118, 1963. 
\bibitem{Caldeira}
Caldeira, A. and Leggett, A. J., Path Integral Approach To Quantum Brownian Motion, Physica, 121A, 587, 1983.
\bibitem{Magpantay}
Magpantay, Jose A., Lindbald Plus From Feynman-Vernon, arxiv:200300194. 
\bibitem{Gradshteyn}
Gradshteyn, I. S. and Ryzhik, I. M., Table of Integrals, Series and Products, Academic Press, Inc., 1965.
\end{thebibliography}

\end{document}